\documentclass[english,reprint,aps,prl, superscriptaddress]{revtex4-1}
\usepackage[T1]{fontenc}
\usepackage[latin9]{inputenc}
\usepackage{amsmath}
\usepackage{amssymb}
\usepackage{graphicx}
\newcommand{\Lnorm}[1]{\left\lVert#1\right\rVert_2}
\newcommand{\specnorm}[1]{\left\lVert#1\right\rVert_\mathcal{L}}
\newcommand{\ket}[1]{\left|#1\right>}
\newcommand{\abs}[1]{\left|#1\right|}

\newcommand{\braket}[2]{\left<\left.#1\right|#2\right>}

\makeatletter

\usepackage{babel}

\makeatother

\usepackage{babel}
\usepackage{tikz}
\usetikzlibrary{shapes,arrows}
\usetikzlibrary{arrows.meta}

\tikzstyle{decision} = [diamond, draw, fill=blue!20, 
text width=4.5em, text badly centered, node distance=3cm, inner sep=0pt]
\tikzstyle{block} = [rectangle, draw, fill=blue!30, 
text width=18.5em, text centered, rounded corners, minimum height=4em]
\tikzstyle{line} = [draw, -{Latex[length=0.3cm,width=0.5cm]}, ultra thick]
\tikzstyle{cloud} = [fill=red!50,draw, ellipse, node distance=3cm,
minimum height=2em]
\begin{document}
\title{Controlling Arbitrary Observables in Correlated Many-body Systems}
\date{\today}

\author{Gerard McCaul}
\email{gmccaul@tulane.edu}
\affiliation{Tulane University, New Orleans, LA 70118, USA}

\author{Christopher Orthodoxou}
\email{christopher.orthodoxou@kcl.ac.uk}
\affiliation{Department of Physics, King's College London, Strand, London, WC2R 2LS, U.K.}

\author{Kurt Jacobs} 
\affiliation{U.S. Army Research Laboratory, Computational and Information Sciences Directorate, Adelphi, Maryland 20783, USA} 
\affiliation{Department of Physics, University of Massachusetts at Boston, Boston, MA 02125, USA} 
\affiliation{Hearne Institute for Theoretical Physics, Louisiana State University, Baton Rouge, LA 70803, USA} 

\author{George H. Booth}
\email{george.booth@kcl.ac.uk}
\affiliation{Department of Physics, King's College London, Strand, London, WC2R 2LS, U.K.}

\author{Denys I. Bondar}
\email{dbondar@tulane.edu}
\affiliation{Tulane University, New Orleans, LA 70118, USA}

\begin{abstract}
Here we present an expanded analysis of a model for the manipulation and control of observables in a strongly correlated, many-body system, which was first presented in [McCaul \textit{et al.}, eprint: arXiv:1911.05006]. A field-free, non-linear equation of motion for controlling the expectation value of an essentially arbitrary observable is derived, together with rigorous constraints that determine the limits of controllability. We show that these constraints arise from the physically reasonable assumptions that the system will undergo unitary time evolution, and has enough degrees of freedom for the electrons to be mobile. Furthermore, we give examples of multiple solutions to generating target observable trajectories when the constraints are violated. Ehrenfest theorems are used to further refine the model, and provide a check on the validity of numerical simulations. Finally, the experimental feasibility of implementing the control fields generated by this model is discussed.
\end{abstract}
\maketitle

\section{Introduction}
The study of the control of quantum systems has a rich history \citep{Werschnik2007}, encompassing a diverse array of strategies. This includes both local control \citep{Kosloff1992,Koslofflocalcontrol} and optimal control~\citep{Serban2005,PhysRevLett.106.190501} that steers a system to a final target state using iterative optimisation~\citep{PhysRevA.37.4950, RevModPhys.80.117, Glaser2015}, possibly under additional constraints~\citep{SpectralConstraints, Kosloffoptimalconstraints}. Separate from this is \emph{tracking control}  \citep{PhysRevA.72.023416,PhysRevA.98.043429,PhysRevA.84.022326,Campos2017, doi:10.1063/1.1582847,doi:10.1063/1.477857}, where a physical system is evolved in such a way that a chosen observable conforms to (or ``tracks'')  a pre-selected trajectory. 

Examples of tracking control abound, with applications as diverse as singularity-free tracking of molecular rotors \citep{PhysRevA.98.043429}, optimising dynamics within the density matrix renormalisation group \citep{PhysRevLett.106.190501, PhysRevA.84.022326}, and spectral dynamical mimicry, where a shaped pulse is used to induce an arbitrary desired spectrum in an atomic system \citep{Campos2017}. In a recent paper \citep{companionletter} a model for the tracking control of a many-electron system was presented without derivation. Here, we expand greatly upon that work, in three principal directions.

First, the tracking model used in Ref.\citep{companionletter} is  motivated in Sec.\ref{sec:Tracking}. Starting from general considerations of an  $N$-electron Hamiltonian, a comprehensive derivation of the tracking equation is present. Additionally, in Sec.\ref{sec:TrackingConstraints} we derive the precise constraints on tracking necessary both to avoid singularities and guarantee a unique evolution for the system. A simple example where these constraints are not obeyed and multiple solutions for the tracking field are possible is also provided. 

Given that in tracking control, one recovers the expected observable trajectory by design, a method of verifying that numerical calculations are physically valid is vital. To this end, we detail in Sec.\ref{sec:Ehrenfest} the application of an Ehrenfest theorem to the model as a way both to verify simulations, and remove nonphysical discontinuities from control fields.

Finally, with the purpose of exploring further the experimental requirements of the control protocol, we examine in Sec.\ref{sec:Experimental} the effect of introducing a frequency cut-off for the control field used to create the `driven imposters' detailed in Ref.\citep{companionletter}. We close in Sec.\ref{sec:Discussion} with a discussion of the results and questions for future work. 

\section{Tracking Model \label{sec:Tracking}}

\subsection{Background}
Our goal is to implement a tracking control \citep{doi:10.1063/1.1582847} model for a general $N$-electron system subjected to a laser pulse described by the Hamiltonian (using atomic units)\citep{hagenkleinert2016}
\begin{align}
	\hat{H} &=   \sum_\sigma\int \frac{{\rm d}x}{2}\hat{\psi}^{\dagger}(x) \left[ i \partial_x -A(t) \right]^2 \hat{\psi}(x)  \nonumber \\
		&+  \sum_{\sigma \sigma^\prime} \int \frac{{\rm d}x {\rm d}x'}{2} \hat{\psi}_{\sigma^\prime}^{\dagger}(x')\hat{\psi}_{\sigma}^{\dagger}(x) U(x-x') \hat{\psi}_{\sigma}(x) \hat{\psi}_{\sigma^\prime}(x')
		\label{eq:GeneralHamiltonian}
\end{align}
where $A(t)$ is the field vector potential and the $\hat{\psi}_{\sigma}(x)$ are standard fermionic field operators satisfying  $\left\{\hat{\psi}_{\sigma^\prime}^{\dagger}(x'),\hat{\psi}_{\sigma}(x)\right\}=\delta_{\sigma \sigma^\prime} \delta(x-x^\prime)$ . Ultimately, we wish to calculate the control field $A_T(t)$, such that the trajectory of an expectation $\langle\hat{O}(t)\rangle$  follows some desired function $O_T(t)$  \citep{PhysRevA.72.023416,PhysRevA.98.043429,PhysRevA.84.022326,Campos2017}. For the sake of specificity, here we derive the control field $A_T(t)$ necessary to control the current expectation, but emphasise that an expression can be derived for an arbitrary expectation using the technique described in Sec.\ref{sec:arbobservable}.  We first re-express the model in an explicitly self-adjoint form using
\begin{align}
\hat{\psi}^{\dagger}(x) \left[ i \partial_x -A(t) \right]^2 &\hat{\psi}(x) \notag =\\& \partial_x\left[{\rm  e}^{iA(t)} \hat{\psi}_{\sigma}(x)\right]^{\dagger} \partial_x\left[{\rm  e}^{iA(t)} \hat{\psi}_{\sigma}(x)\right].
\end{align}
In this form, one may straightforwardly construct a continuity equation for the density operator $\hat{\rho}(x) = \hat{\psi}^{\dagger}(x) \hat{\psi}(x)$:
\begin{align}\label{Eq:Continuity} 
	\frac{{\rm d}}{{\rm d}t} \hat{\rho}(x) = i \left[\hat{H}, \hat{\rho}(x)\right] 
		= -\partial_x  \hat{J}(x),
\end{align}
which defines the current operator $\hat{J}(x)$,
\begin{align}
	\hat{J}(x) =& \frac{1}{2i}\left[ \hat{\psi}^{\dagger}(x)\partial_x \hat{\psi}(x) - \partial_x \hat{\psi}^{\dagger}(x) \hat{\psi}(x) \right] \notag \\ &+ A(t) \hat{\psi}^{\dagger}(x) \hat{\psi}(x)  . \label{eq:currentcontinuum}
\end{align}
The current expectation is obtained from this expression by taking expectations and integrating over space, i.e. $\int {\rm d}x\left< \hat{J}(x)\right>=J(t)$. Noting that $N = \left\langle \int \hat{\rho}(x) dx \right\rangle$ is a conserved quantity, one may straightforwardly invert Eq.\eqref{eq:currentcontinuum} to obtain the $A_T(t)$ that corresponds to $\int {\rm d}x\left< \hat{J}(x)\right>=J_T(t)$:
\begin{align}\label{Eq:EhrenfestCurrent}
	A_T(t) =& \frac{i}{2N} \int dx \left\langle \hat{\psi}^{\dagger}(x)\partial_x \hat{\psi}(x) - \partial_x \hat{\psi}^{\dagger}(x) \hat{\psi}(x) \right\rangle(t) \notag \\ &+ \frac{J_T(t)}{N}.
\end{align}
For systems with Bosonic statistics, it is easy to show that the control field equation is almost identical, but the definition of the current operator picks up a negative sign $\hat{J}(x)\to -\hat{J}(x)$.

\subsection{Tracking Control in A Discrete Model}
While the equation for the tracking control field will in principle describe tracking for an $N$-electron system, in this paper we will provide a concrete illustration of its use with a lattice model. To do so, we first discretise the model Hamiltonian, using $a$ as the lattice constant such that $x=ja$ and $x^\prime =ka$:
\begin{align}
	\int dx &\to \sum_{r} a \Longrightarrow 
	\delta(x-x') \to \frac{\delta_{jk}}{a}, \\ 
	\hat{\psi}_{\sigma}(x) &\to \frac{\hat{c}_{j\sigma}}{\sqrt{a}} \Longrightarrow
	\left\{ \hat{c}_{j\sigma}^{\dagger},  \hat{c}_{k\sigma'} \right\} = \delta_{jk} \delta_{\sigma \sigma'}, \\
	\partial_x g(x) &\to [g_{j+1} - g_j]/a.
\end{align}
After discretisation and assuming periodic boundary conditions, the Hamiltonian takes the form: 
\begin{align}
	\hat{H} =& -\sum_{j,\sigma} \frac{1}{2a} \left( e^{-i\Phi(t)} \hat{c}_{j+1 \sigma}^{\dagger} \hat{c}_{j\sigma} + e^{i\Phi(t)} \hat{c}_{j \sigma}^{\dagger} \hat{c}_{j+1, \sigma}  \right) \notag\\
		& +\sum_{j,\sigma} \frac{1}{a}
		\hat{c}_{j\sigma}^{\dagger} \hat{c}_{j \sigma}+ \sum_{j,k,\sigma,\sigma'} \frac{1}{2a^2} U_{j-k} \hat{c}_{k\sigma'}^{\dagger} \hat{c}_{k \sigma}^{\dagger} \hat{c}_{j\sigma} \hat{c}_{j\sigma'}
\end{align}
where we have set $\Phi(t)=aA(t)$. From this discretised Hamiltonian, one is able to derive a continuity equation for $\hat{\rho_j}=\sum_{ \sigma}\hat{c}_{j\sigma}^{\dagger} \hat{c}_{j \sigma}$:
\begin{align}
    \frac{{\rm d}\hat{\rho_j}}{{\rm d}t}&=\frac{1}{a}(\hat{J}_j-\hat{J}_{j-1}),  \\
    \hat{J}_j&=-i\sum_{\sigma}\left({\rm e}^{-i\Phi\left(t\right)}\hat{c}_{j\sigma}^{\dagger}\hat{c}_{j+1\sigma}-{\rm h.c.}\right).
\end{align}
This continuity equation defines the current operator $\hat{J}=\sum_j \hat{J}_j$, and has the important property of being composed only from the kinetic part of the Hamiltonian. This means the current operator is not explicitly dependent on the form of the interaction $U_{j-k}$. As a result of this property, the construction of a method to track the expectation of the current operator does not depend on the specific form of the Hamiltonian's interparticle interactions. For this reason, we will restrict our derivation to a specific Hamiltonian, but emphasise that the results may be applied to any model with the form of  Eq.(\ref{eq:GeneralHamiltonian}).

\begin{figure}
\begin{center}
\includegraphics[width=0.8\columnwidth]{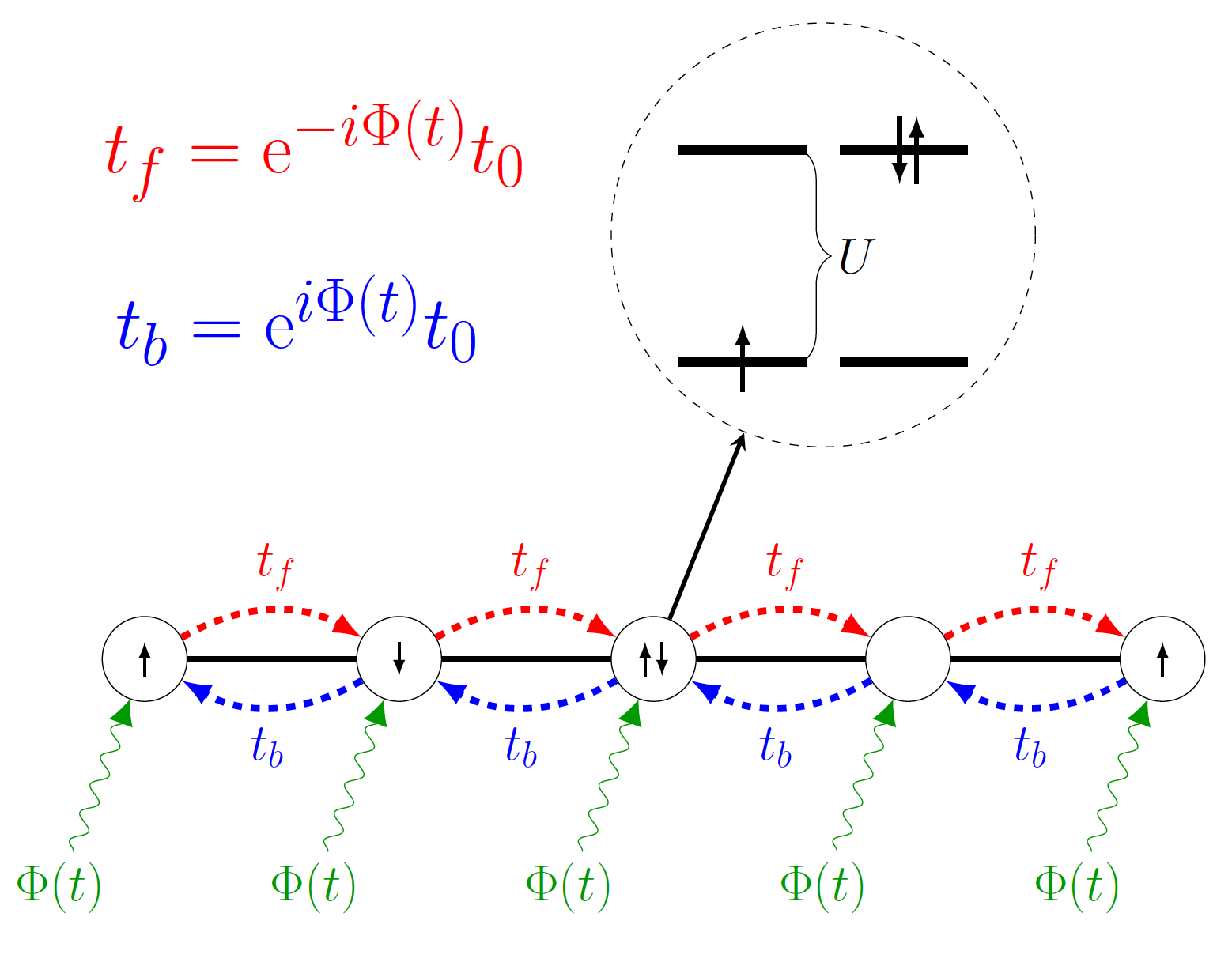}\end{center}\caption{Schematic representation of the Fermi-Hubbard model.  Electrons hop between sites with an on-site repulsion of $U$, and a hermitian hopping amplitude scaled by the applied field $\Phi(t)$.}
\label{fig:Hamiltonian}
\end{figure}
From this point forward we will use the 1D Fermi-Hubbard model \citep{Tasaki1998} (see Fig.~\ref{fig:Hamiltonian} for a schematic representation) as a concrete example of the tracking strategy. This model has the Hamiltonian
\begin{align}
\hat{H}\text{\ensuremath{\left(t\right)}}= & -t_{0}\sum_{j,\sigma}\text{\ensuremath{\left({\rm e}^{-i\Phi\left(t\right)}\hat{c}_{j\sigma}^{\dagger}\hat{c}_{j+1\sigma}+{\rm e}^{i\Phi\left(t\right)}\hat{c}_{j+1\sigma}^{\dagger}\hat{c}_{j\sigma}\right)}}\nonumber \\
 & +U\sum_{j}\hat{c}_{j\uparrow}^{\dagger}\hat{c}_{j\uparrow}\hat{c}_{j\downarrow}^{\dagger}\hat{c}_{j\downarrow}.
\label{eq:Hamiltonian}\end{align}
As in the continuum case, we wish to find the vector potential that will produce a specified current $J_T\left(t\right)=\left\langle \hat{J}\right\rangle $. To do so, we take the current expectation

\begin{equation}
\hat{J}=-iat_{0}\sum_{j,\sigma}\left({\rm e}^{-i\Phi\left(t\right)}\hat{c}_{j\sigma}^{\dagger}\hat{c}_{j+1\sigma}-{\rm h.c.}\right),\label{eq:currentoperator}
\end{equation}
and rearrange for $\Phi$, expressing
the nearest neighbour expectation in a polar form:
\begin{equation}
\left\langle \psi (t) \left|\sum_{j,\sigma} \hat{c}_{j\sigma}^{\dagger}\hat{c}_{j+1\sigma}\right| \psi (t) \right\rangle =R\left(\psi\right){\rm e}^{i\theta\left(\psi\right)}. \label{neighbourexpectation}
\end{equation}
In both Eq.\eqref{neighbourexpectation} and later expressions, the  argument $\psi$ indicates that the expression is dependent on a functional of $\ket{\psi}\equiv \ket{\psi(t)}$. Eq.\eqref{neighbourexpectation} can be used in conjunction with Eq.\eqref{eq:currentoperator} to yield
\begin{align}
J\left(t\right)= & -i a t_{0} R\left(\psi\right)\left({\rm e}^{-i\left[\Phi\left(t\right)-\theta\left(t\right)\right]}-{\rm e}^{i\left[\Phi\left(t\right)-\theta\left(\psi\right)\right]}\right)\nonumber \\
= & -2 a t_{0} R \left(\psi\right)\sin(\Phi\left(t\right)-\theta\left(\psi\right)).\label{eq:currentexpectation}
\end{align}
An important caveat that should be noted here is that if one were to apply a time dependent rotation to the system, the current expectation would no longer  depend explicitly on $\Phi(t)$ \citep{PhysRevA.95.023601}, but instead there would remain an implicit dependence through the state of the system $\ket{\psi}$. This is important, as in order to define a control field which reproduces a tracking current $J_T(t)$, we invert Eq.(\ref{eq:currentexpectation}). From this inversion we obtain the tracking control field $\Phi_T(t, \psi)$, which takes the desired current expectation as a parameter,
\begin{equation}
\Phi_T\left(t, \psi \right)=\arcsin\left[-X(t,\psi)\right]+\theta\left(\psi\right).
\label{eq:phi_track}
\end{equation}
in which we have defined 
\begin{align}
   X(t,\psi) & = \frac{J_T\left(t\right)}{2at_{0}R\left(\psi\right)}  . 
\end{align}
From Eq.(\ref{eq:phi_track}) it is possible to eliminate the control field entirely from the model Hamiltonian using the equality
\begin{align}
{\rm e}^{\pm i\Phi_{T}\left(t,\psi\right)}={\rm e}^{\pm i\theta\left(\psi\right)}\left[\sqrt{1- X^2(t,\psi)}\mp iX(t,\psi)\right],
\end{align}
where the above equality is obtained via Euler's equation and $\cos\left(\arcsin\left(x\right)\right)=\sqrt{1-x^{2}}$.  From this, we are able to define the ``tracking Hamiltonian'' $\hat{H}_T (J_T(t),\psi)$ which takes the target current $J_T(t)$ as a parameter:
\begin{align}
\hat{H}_{T}\left(J_T(t), \psi\right) & = \sum_{\sigma,j} \left[ P_+{\rm e}^{-i\theta\left(\psi \right)}\hat{c}_{j\sigma}^{\dagger}\hat{c}_{j+1\sigma} + \mbox{H.c.} \right] \nonumber \\
 & \;\;\;\; + U\sum_{j}\hat{c}_{j\uparrow}^{\dagger}\hat{c}_{j\uparrow}\hat{c}_{j\downarrow}^{\dagger}\hat{c}_{j\downarrow}, \label{eq:trackingHamiltonian} \\
P_{\pm} & = -t_{0}\left(\sqrt{1 - X^2(t,\psi)}\pm i X(t,\psi)\right).
\end{align}
This leads to a field-free, non-linear evolution for the wavefunction given by
\begin{align}
    i\frac{{\rm d}\left|\psi\right>}{{\rm d}t}=\hat{H}_T\left(J_T(t), \psi\right)\left|\psi\right>, \label{eq:eqnofmotion}
\end{align}
which is equivalent to evolving the system with the original Hamiltonian given in Eq.(\ref{eq:Hamiltonian})  and the usual Schr{\"o}dinger equation $i\frac{{\rm d}\left|\psi\right>}{{\rm d}t}=\hat{H}\left(t\right)\left|\psi\right>$, under the additional constraint that $\Phi(t)$ is chosen such that $\langle \hat{J}(t)\rangle =J_T (t)$. After solving Eq.\eqref{eq:eqnofmotion}, it is also possible to recover the tracking field $\Phi_T(t)$ via Eq.\eqref{eq:phi_track}.

\subsection{Tracking Arbitrary Observables \label{sec:arbobservable}}
Finally, we extend the derivation for tracking current to an arbitrary observable $\hat{O}=\hat{O}^\dagger$ whose expectation $O(t)=\left\langle \hat{O}\right\rangle$  is not a function of $\Phi$. In this case the time derivative is
\begin{align}
  \frac{{\rm d}O(t)}{{\rm d} t}=&it_{0}\sum_{j,\sigma}\text{\ensuremath{\left({\rm e}^{-i\Phi\left(t\right)}\left\langle \left[\hat{c}_{j\sigma}^{\dagger}\hat{c}_{j+1,\sigma},\hat{O}\right]\right\rangle +{\rm h.c.} \right)}}
\nonumber \\ &-iU\sum_{j}\left\langle \left[\hat{c}_{j\uparrow}^{\dagger}\hat{c}_{j\uparrow}\hat{c}_{j\downarrow}^{\dagger}\hat{c}_{j\downarrow},\hat{O}\right]\right\rangle .
\end{align}
From this evolution, we assign
\begin{align}
\sum_{j,\sigma}\left\langle \left[\hat{c}_{j\sigma}^{\dagger}\hat{c}_{j+1,\sigma},\hat{O}\right]\right\rangle =R_{O}{\rm e}^{i\theta_{O}}, \\
B=-iU\sum_{j}\left\langle \left[\hat{c}_{j\uparrow}^{\dagger}\hat{c}_{j\uparrow}\hat{c}_{j\downarrow}^{\dagger}\hat{c}_{j\downarrow},\hat{O}\right]\right\rangle.
\end{align}
With this substitution we obtain an expression for the derivative of the observable in terms of the control field:
 \begin{align}
  \frac{{\rm d}O(t)}{{\rm d} t}=-2t_{0}R_{O}\sin\left(\Phi-\theta_{O}\right)+B.
  \end{align}
This can be inverted to obtain the tracking control field for an arbitrary observable 
   \begin{align}
   \Phi_{O}=\arcsin\left(\frac{B-\frac{{\rm d}O}{{\rm d} t}}{2t_{0}R_{O}}\right)+\theta_{O}.
   \end{align}
  From this a tracking Hamiltonian and constraint can be derived using the methods presented previously. The theoretical considerations in the rest of this paper may be applied to tracking an arbitrary variable, but in the interests of clarity we shall restrict our attention to tracking of the current expectation using Eq.(\ref{eq:trackingHamiltonian}). 

\section{Tracking Constraints \label{sec:TrackingConstraints}}
In this section we prove the statement:

\emph{For a finite system, if the wavefunction $\ket{\psi}\equiv\ket{\psi(t)}$ solves Eq.\eqref{eq:eqnofmotion}, and satisfies the constraints}
\begin{align}
\left|X(t, \psi)\right|&<1-\epsilon_1 \label{eq:constraint2}, \\
R(\psi)&>\epsilon_2, \label{eq:constraint1}
\end{align}
\emph{where $\epsilon_1, \epsilon_2$ are any positive constants, then $\ket{\psi}$ is a unique solution of Eq.\eqref{eq:eqnofmotion} and therefore by Eq.\eqref{eq:phi_track}, $\Phi_T(t)$ is a unique field which solves the current tracking problem.}

Both the constraints given by Eqs.(\ref{eq:constraint2},\ref{eq:constraint1}) are necessary conditions for $\hat{H}_T\left(J_T(t), \psi\right)$ to be \emph{Lipschitz continuous} (LC) over $\ket{\psi}$ \citep{geraldfolland2007}. In this case, the Picard-Lindel{\"o}f theorem guarantees $\ket{\psi}$ has a unique solution depending on its initial value when being evolved by the tracking Hamiltonian \citep{kentnagle2011}. 

In Sec.\ref{sec:formalproof}, we show formally that under the constraints given by Eqs.(\ref{eq:constraint2},\ref{eq:constraint1}), the tracking Hamiltonian is LC, while in Sec.\ref{sec:motivation} we provide a physical motivation for these constraints. Finally in Sec.\ref{sec:multiple} we provide a simple example where the derived constraints do not hold, and multiple solutions for the tracking field are possible. 

\subsection{Proving Lipschitz continuity \label{sec:formalproof}}
We define the $L_2$ norm $\Lnorm{\ket{\psi}}=\sqrt{\braket{\psi}{\psi}}$ and spectral norm \citep{horn1985matrix}:
\begin{equation}
    \specnorm{\hat{A}}=\sup_{\braket{\psi}{\psi}=1}\Lnorm{\hat{A}\ket{\psi}}.
\end{equation}
These norms obey a submultiplicative property \citep{borzi},
\begin{equation}
\Lnorm{\hat{A}\ket{\psi}}\leq\specnorm{\hat{A}}\Lnorm{\ket{\psi}}
\end{equation}
which when combined with the Cauchy-Schwarz inequality yields:
\begin{equation}
\left|\left\langle \phi\left|\hat{A}\right|\psi \right\rangle\right|\leq \Lnorm{\ket{\phi}}\Lnorm{\hat{A}\ket{\psi}}\leq\Lnorm{\ket{\phi}}\specnorm{\hat{A}}\Lnorm{\ket{\psi}}. \label{eq:Cauchy}
\end{equation}

We now proceed to proving that for the set of wavefunctions which obey Eqs.(\ref{eq:constraint2}, \ref{eq:constraint1}), the following inequality holds:
\begin{equation}
\Lnorm{\hat{H}_T\left(J_T(t), \psi\right)\ket{\psi}- \hat{H}_T\left(J_T(t), \phi\right)\ket{\phi}} \leq L_H \Lnorm{\ket{\psi}-\ket{\phi}} \label{eq:TrackingLipschitz}
\end{equation}
where $L_H$ is some finite constant, and is the definition of LC for the function $\hat{H}_T\left(J_T(t), \psi\right)\ket{\psi}$. In order to prove this, it is convenient to establish some properties both for operators and functionals of $\ket{\psi}$.

First, in finite dimensions all linear operators are bounded, which implies they are also LC over the whole Hilbert space:
\begin{equation}
\Lnorm{\hat{A}\left(\ket{\psi}-\ket{\phi}\right)}\leq \specnorm{\hat{A}}\Lnorm{\ket{\psi}-\ket{\phi}}.
\end{equation}
Additionally, the expectation of linear operators $\left\langle \psi \left|\hat{A} \right| \psi \right\rangle =A(\psi)$ is also LC on the space of wavefunctions ($\Lnorm{\psi}=1$). This is demonstrated by taking the identity
\begin{align}
A(\psi)-&A(\phi)=\left\langle\psi\left|\hat{A}\right|\psi\right\rangle- \left\langle\phi\left|\hat{A}\right|\phi\right\rangle  \notag\\  &=\left\langle\psi\right|\hat{A}\left( \left|\psi\right\rangle-\left|\phi \right\rangle\right)- \left(\left\langle\phi \right|-\left\langle\psi\right| \right)\hat{A}\left|\phi\right\rangle,
\end{align}
and applying the triangle inequality  $\abs{x+y}\leq\abs{x}+\abs{y}$ to its norm:
\begin{align}
\abs{A(\psi)-A(\phi)} \leq2\specnorm{\hat{A}} \Lnorm{\left|\psi \right\rangle-\left|\phi \right\rangle}. \label{eq:expectationinequality}
\end{align}

More generally, an arbitrary functional of $\ket{\psi}$, $f:\ket{\psi}\to\mathbb{C}$  is LC over $\ket{\psi}$ if for all $\ket{\psi},\ket{\phi}$ in its domain, it satisfies the inequality 

\begin{equation}
\ensuremath{\abs{f\left(\psi\right)-f\left(\phi\right)}\leq}L_f\Lnorm{\ket{\psi}-\ket{\phi}}
\end{equation}
where $L_f$ is some finite constant. Taking two functionals $f\left(\psi\right)$, $g\left(\psi\right)$, which are LC over $\ket{\psi}$ with Lipschitz constants $L_f$ and $L_g$, then the norm of their product $h\left(\psi\right)=f\left(\psi\right)g\left(\psi\right)$
is:
\begin{align}
\abs{h(\psi)-h(\phi)}  &=\abs{\left(f(\psi)-f(\phi)\right)g(\psi)+f(\phi)\left(g(\psi)-g(\phi)\right)}\nonumber \\
  &\leq\abs{f(\psi)-f(\phi)}\abs{g(\psi)}+\abs{f(\phi)}\abs{g(\psi)-g(\phi)} \nonumber \\
   &\leq  \left( L_f\abs{g(\psi)}+L_g\abs{f(\phi)} \right)\ensuremath{\Lnorm{\ket{\psi}-\ket{\phi}}} \label{eq:Lipschitzproductscalar}.
\end{align}
This means that if the functionals $f(\psi)$, $g(\psi)$ are LC \emph{and} bounded over the domain of $\psi$ then their product is also LC. In the case of a product between an operator and an LC functional, $f(\psi)\hat{A}$, a similar result to Eq.\eqref{eq:Lipschitzproductscalar} is obtained:
\begin{align}
\Lnorm{f(\psi)\hat{A}\ket{\psi}-f(\phi)\hat{A}\ket{\phi}} &\notag \\ \leq  \specnorm{\hat{A}}&\left( L_f+\abs{f(\psi)} \right)\ensuremath{\Lnorm{\ket{\psi}-\ket{\phi}}} \label{eq:Lipschitzproductmix},
\end{align}
i.e. if $f(\psi)$ is bounded and LC, $f(\psi)\hat{A}$ is also LC. Lastly, sums of any LC operators or functionals will themselves be LC by the triangle inequality. 

Equipped with these properties, the most direct route to proving Eq.\eqref{eq:TrackingLipschitz} is to prove each of the constituent components of Eq.(\ref{eq:trackingHamiltonian}) are both LC and bounded, which by Eqs.(\ref{eq:Lipschitzproductscalar}, \ref{eq:Lipschitzproductmix})  and the triangle inequality is sufficient to prove that the tracking Hamiltonian is itself LC in $\psi$. 

The relevant parts of the Hamiltonian for which Lipschitz continuity over $\ket{\psi}$ and boundedness must be demonstrated are $\rm{e}^{ i\theta(\psi)}$ and $P_\pm$. To prove the former is LC, we first consider the nearest neighbour expectation, using $\sum_{j,\sigma}\hat{c}_{j\sigma}^{\dagger}\hat{c}_{j+1\sigma}=\hat{K}$:
\begin{equation}
\left\langle \psi \left|\hat{K} \right| \psi \right\rangle=K(\psi)=R(\psi){\rm e}^{i\theta(\psi)}. \label{eq:Kneighbour}
\end{equation}
This expectation is LC by Eq.\eqref{eq:expectationinequality}, and bounded due to Eq.\eqref{eq:Cauchy} and the normalisation of wavefunctions.
Combining this result with the reverse triangle inequality $\big|  \left| x\right| - \left|y \right| \big| \leq \left| x-y \right|$ further demonstrates that $R(\psi)=\left|K(\psi)\right|$ is also LC and bounded. The final step in order to show ${\rm e}^{i\theta(\psi)}$ is itself LC is to establish $R^{-1}(\psi)$ is LC under Eqs.(\ref{eq:constraint2}, \ref{eq:constraint1}). This is easily established by
\begin{align}
\abs{R^{-1}(\psi)-R^{-1}(\phi)}=& \frac{1}{R(\psi)R(\phi)}\abs{R(\psi)-R(\phi)} \notag \\ \leq& \frac{1}{\epsilon^2_2}\specnorm{\hat{K}}\Lnorm{\left|\psi \right\rangle-\left|\phi \right\rangle}
\end{align}
where in the second inequality we have utilised Eq.\eqref{eq:constraint1}. By Eq.(\ref{eq:Lipschitzproductscalar}) we therefore establish 
${\rm e}^{i\theta(\psi)}=\frac{K(\psi)}{R(\psi)}$ is LC, and is bounded by definition.

The final term to tackle is $P_\pm$. Since this is the only term that involves our target $J_T(t)$, we work directly in the variable $x=X(t,\psi)$. The function $f(x)=x$ is itself trivially LC and bounded over this domain where Eq.\eqref{eq:constraint2} is satisfied. It therefore only remains to check the Lipschitz continuity of $f(x)=\sqrt{1-x^2}$. Since this function is differentiable on the interval $I=[-(1-\epsilon_2),1-\epsilon_2]$ which satisfies Eq.\eqref{eq:constraint2}, by the mean value theorem \citep{doi:10.1112/plms/s2-7.1.14} the function is LC if $\left|f^\prime(x)\right|\leq M$ for all $x \in I$ and $M$ is finite. It is easy to show that
\begin{equation}
M=\max_{x\in I}\left|f^\prime(x)\right|=\frac{1-\epsilon_2}{2\sqrt{2\epsilon_2-\epsilon_2^2}}
\end{equation}
and therefore $P_\pm$ is LC and bounded provided $\epsilon_2\neq0$ and $<1$ . As a result, we establish that under the conditions of Eqs.(\ref{eq:constraint2}, \ref{eq:constraint1}), each of the components of the tracking Hamiltonian is LC and bounded, meaning that the Hamiltonian is itself LC. From this continuity it follows that the Picard-Lindel{\"o}f theorem is obeyed and $\ket{\psi}$ has a unique solution depending on its initial value. It is interesting to note that this result, derived from the analysis of the continuous formulation \eqref{eq:trackingHamiltonian}, stands in sharp contrast to some discretized approaches to tracking problems, in which multiple solutions are possible \citep{Rabitzmultiple}.

\subsection{Physical Motivation \label{sec:motivation}} 
It is reasonable to ask whether the constraints imposed upon $\psi$ are well justified, and here we provide physical motivation for them. First, the condition $\left|X(t, \psi)\right|<1-\epsilon_1$ is easily justified by noting that if this is violated, $P_+^\dagger \neq P_-$ and the tracking Hamiltonian in Eq.\eqref{eq:trackingHamiltonian} is no longer Hermitian. This constraint therefore corresponds to a restriction on the currents that can be produced in a physical system to ensure that the state undergoes appropriate unitary time evolution.

\begin{figure}
\begin{center}
\includegraphics[width=0.8\columnwidth]{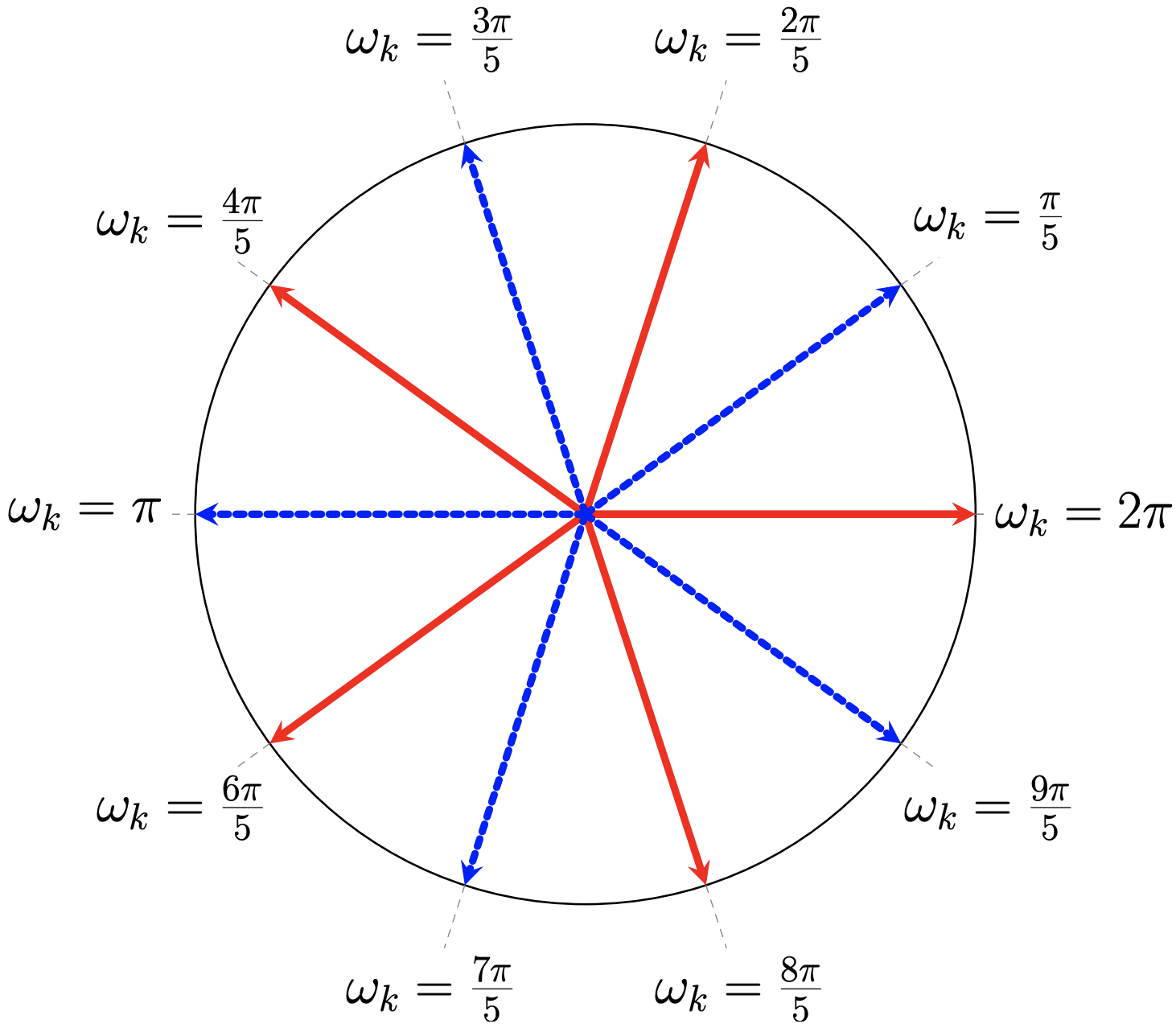}\end{center}\caption{An example of the contributions to $K(\psi)$ for $L=10$ sites at half filling. In this example the occupied states for one spin species (dashed blue) have been chosen so that they are in anti-phase with the other species (red), and therefore $K(\psi)=0$.}
\label{fig:components}
\end{figure}

The restriction imposed by Eq.\eqref{eq:constraint1} is somewhat more general as it does not make reference to the current being tracked. Nevertheless, we shall demonstrate here that it is reasonable to expect this property in physical systems. We first consider $\hat{K}$ in a diagonal basis, using the transformation $\hat{c}_{j\sigma}=\sum_{k}{\rm e}^{i\omega_{k}j}\tilde{c}_{k\sigma}$
 where $\omega_{k}=\frac{2\pi k}{L}$, and $L$ is the number of sites. The nearest neighbour expectation then assumes the
form
\begin{equation}
K(\psi) =\sum_{k,\sigma}\left(\cos\left(\omega_{k}\right)+i\sin\left(\omega_{k}\right)\right)\left\langle \psi\left|\tilde{c}_{k\sigma}^{\dagger}\tilde{c}_{k\sigma}\right|\psi\right\rangle. 
\end{equation} 
In the diagonal space, we immediately see that every occupied state in momentum space contributes components with equal magnitude but which differ by a phase. For an even number of particles (as is always the case at half filling), it is mathematically very easy to construct an arbitrary wavefunction such that each occupied state's contribution is in antiphase with another, making $K(\psi)=0$ and violating the tracking constraint. A simple example of this is shown in Fig.\ref{fig:components}. 
 
 While it is possible to construct a wavefunction which violates Eq.\eqref{eq:constraint1}, the question is whether such a wavefunction is truly physical. To answer this, we consider Eq.(\ref{eq:Hamiltonian}) in the tight binding limit ($\frac{U}{t_0}=0$). In the diagonalised basis, this Hamiltonian is \citep{floriangebhard2010,fabianessler2005}

\begin{equation}
\hat{H}\left(t\right)=-2t_{0}\sum_{k,\sigma}\cos(\omega_k-\Phi(t))\tilde{c}_{k\sigma}^{\dagger}\tilde{c}_{k\sigma}.
\end{equation}
 Notice that this shares a common eigenbasis with the nearest neighbour
expectation. Before any driving occurs, the system is in the ground state $\left|\psi_{g}\right>$,
that minimises the system energy. Since the Hamiltonian is diagonal in the occupation number basis, the ground state will be a pure state \footnote{In fact, the ground state is potentially highly degenerate depending on the number of sites and filling fraction, but since by we are only arguing that the ground state energy is non-zero, it is sufficient to treat the ground state as a pure state in the occupation number representation.} in this representation, and has energy:
\begin{equation}
\left\langle \psi_{g}\left|\hat{H}\left(0\right)\right|\psi_{g}\right\rangle =-2t_{0}\sum_{k,\sigma}\cos\left(\omega_{k}\right)\delta(k,\sigma)=E_{g} \label{eq:groundstate}
\end{equation}
 where $\delta(k,\sigma)=\left\langle {\psi}_{g}\left|\tilde{c}_{k\sigma}^{\dagger}\tilde{c}_{k\sigma}\right|{\psi}_{g}\right\rangle$ is $1$ if the relevant mode is occupied in the ground state, and zero otherwise. 
 
 Clearly, the occupation numbers of the ground state will be such that Eq.(\ref{eq:groundstate}) is \emph{minimised}. If one has $N=\sum_\sigma N_\sigma$ particles on an $L$ site lattice, each spin species' contribution to the ground state energy will consist of the $N_\sigma$ momentum modes closest to $\omega_k$=0. From this counting argument, it is possible to give an analytic expression for $E_g=\sum_\sigma E_\sigma$:
 \begin{equation}
    -\frac{E_\sigma}{2t_0} = \left\{
        \begin{array}{ll}
        1+2\sum\limits^{\frac{N_\sigma}{2}-1}_{k=1}\cos\omega_{k} + \cos\frac{\pi N_\sigma}{L} & \textrm{if $N_\sigma >0$ is even}, \\
        1+2\sum\limits^{\frac{N_\sigma-1}{2}}_{k=1}\cos\omega_{k}  & \textrm{if $N_\sigma$ is odd}.
        \end{array}
    \right.
\end{equation}
It is easy to see from this analytic expression that the only cases for which $E_g$ is zero is either the vacuum or when every mode of both spin species is occupied, and the system dynamics are completely frozen.

Having established $E_g$ is non-zero in all but the most trivial of circumstances, we now substitute it into the nearest neighbour expectation to obtain $K(\psi_g)$:
\begin{align}
K(\psi_g)&=\left\langle \psi_{g}\left|\sum_{j,\sigma}\hat{c}_{j\sigma}^{\dagger}\hat{c}_{j+1\sigma}\right|\psi_{g}\right\rangle =-\frac{E_{g}}{2t_{0}}+\lambda, \\
\lambda&=i\sum_{k}\delta(k,\sigma)\sin\left(\omega_{k}\right)
\end{align}
 which means that for the ground state, $K(\psi_g)$ has
a non-zero real part and $R\left(\psi_g\right)$ \emph{must }be non-zero. Furthermore, since the Hamiltonian and
nearest neighbour operators commute at all times in the diagonal basis, the value of $R\left(\psi\right)$ is time-independent and therefore non-zero for all $\psi$ that can be evolved from the ground state. 

In a system with non-zero $U$, we can consider only the kinetic term, which has the form
\begin{equation}
\hat{H}_{K}=-t_{0}\sum_{j,\sigma}\left(\hat{c}_{j\sigma}^{\dagger}\hat{c}_{j+1\sigma}+\hat{c}_{j+1\sigma}^{\dagger}\hat{c}_{j\sigma}\right).
\end{equation}
In this case, provided $\left\langle \psi\left|\hat{H}_K\right|\psi\right\rangle\neq0$, an analogous argument can be made to justify Eq.\eqref{eq:constraint1}. For this reason, we
can consider that this constraint corresponds to the condition
that there is some kinetic energy in the system, and the electrons have not been completely frozen (a natural precondition for observing \emph{any} current). 
 
 We conclude this section with the observation that while in principle the derived constraints are highly non-linear inequalities in $\ket{\psi}$, in practice simulations confirm the expectation that even at high $\frac{U}{t_0}$, Eq.\eqref{eq:constraint1} is obeyed (see e.g. Fig.\ref{fig:Ehrenfesttracking}). Furthermore, it is relatively easy to satisfy Eq.\eqref{eq:constraint2} via a heuristic scaling of the target to be tracked, as these constraints limit only the peak amplitude of current in the evolution, and otherwise allow for any function to be tracked when appropriately scaled. If one is concerned only with reproducing the shape of the target current, then using a scaled target $J_s(t)=kJ_T(t)$ such that $\left|J_s\left(t\right)\right|<2at_{0}R\left(t\right)$  will allow tracking unproblematically. Alternately, if one treats the lattice constant $a$ as a tunable parameter, this can always be set for the tracking system so as to satisfy $\left|X(t, \psi)\right|<1-\epsilon_2$.

 Singularities in the control field are a common occurrence in tracking control, which often make a specified trajectory impossible to reproduce \citep{doi:10.1063/1.1582847, doi:10.1137/0325030, PhysRevA.98.043429}. While singularities are present in the unconstrained model presented here, they are easily identified and avoided using the constraints derived above.  

\subsection{Multiple Solutions \label{sec:multiple}} 
We conclude this section with a demonstration that when the derived constraints of Eqs.(\ref{eq:constraint2},\ref{eq:constraint1})  are not both satisfied, multiple solutions for $\ket{\psi}$ and hence $\Phi_T(t)$ are possible. To simplify algebra, we consider a $U=0$ system, where $\theta(\psi)=0$ regardless of the field applied when evolving from the ground state.

Now consider a situation where one uses  tracking simply to reproduce the current produced by some field $\Phi(t)$, i.e. $J_T(t)=J(t)$ and if the solution is unique, $\Phi_T=\Phi(t)$. Applying tracking to this situation, if
\begin{equation}
    \label{eq:multiiplecondition}
    \Phi(0)=0, \qquad
    \abs{\Phi(t)} < \frac{\pi}{2},
\end{equation}
then the solution is unique and  $\Phi_T(t)=\Phi(t)$.

\begin{figure}
\begin{center}
\includegraphics[width=1.1\columnwidth]{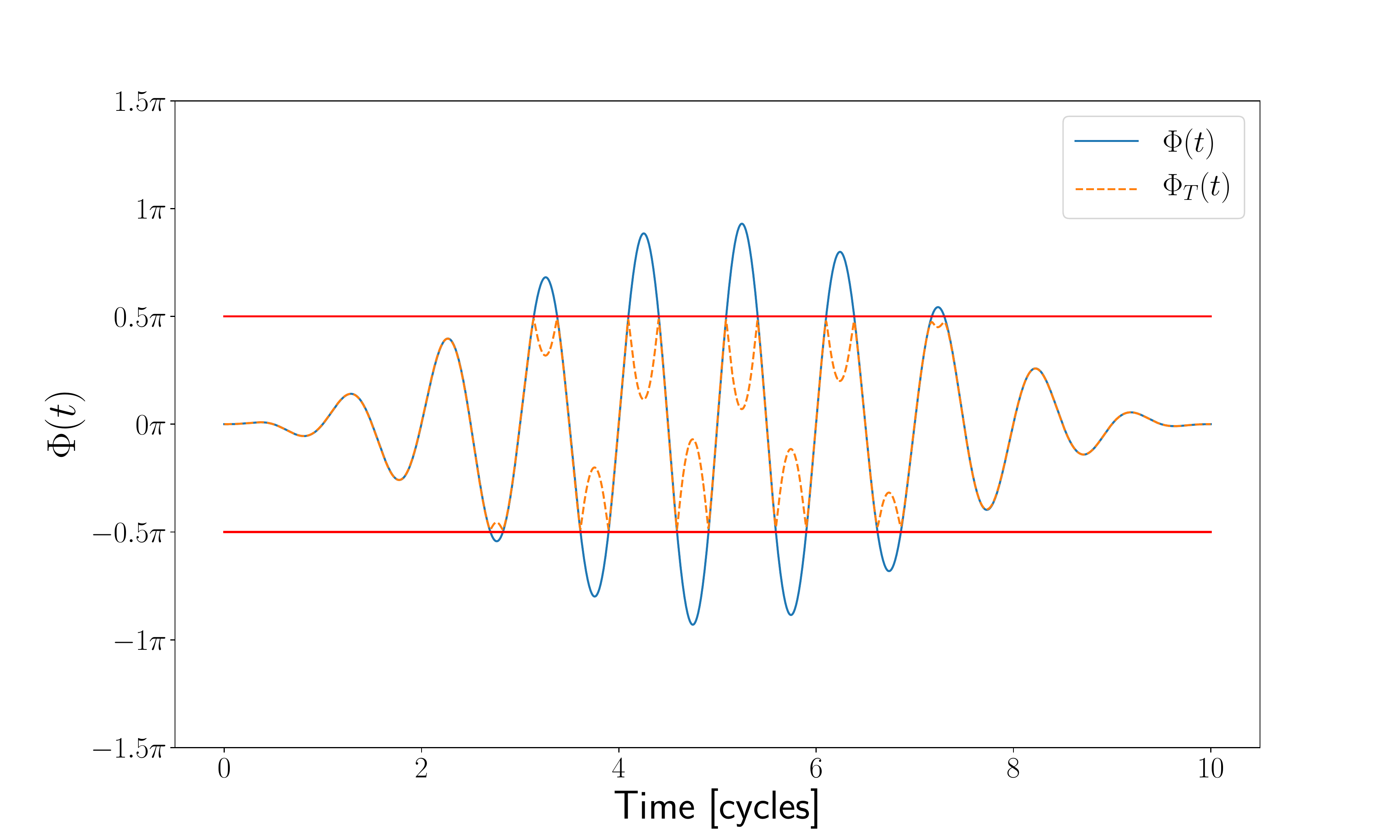}\end{center}\caption{Control fields driving a $U=0$ system, each of which generates the same current. Here there are multiple solutions, $\Phi(t)\neq\Phi_T(t)$ due to the violation of Eq.\eqref{eq:constraint2} at $\Phi(t)=\pm\frac{\pi}{2}$.}
\label{fig:multiplesimulation}
\end{figure}

If however there is a point where $\abs{\Phi(t)} = \frac{\pi}{2}$ then  $X(t,\psi)= 1$ and Eq.\eqref{eq:constraint2} is violated. If the control field is continuous, then any $\Phi(t)$ which does not obey Eq.\eqref{eq:multiiplecondition} also violates Eq.\eqref{eq:constraint2}.  Fig.\ref{fig:multiplesimulation} confirms this violation, where both control fields generate the same current (shown in Fig.\ref{fig:refEhrenfest}\textbf{(a)}), but have different functional forms. 

The multiplicity of solutions shown can be understood physically in a simple manner. Reproducing the target current  only requires that $\sin(\Phi(t))=\sin(\Phi_T(t))$, but identical dynamics requires ${\rm e}^{\pm i\Phi(t)}={\rm e}^{\pm i\Phi_T(t)}$. The latter condition is much stricter, and only coincides with the tracking requirements when Eq.\eqref{eq:multiiplecondition} is also obeyed. 
\begin{figure}
\begin{center}
\includegraphics[width=0.8\columnwidth]{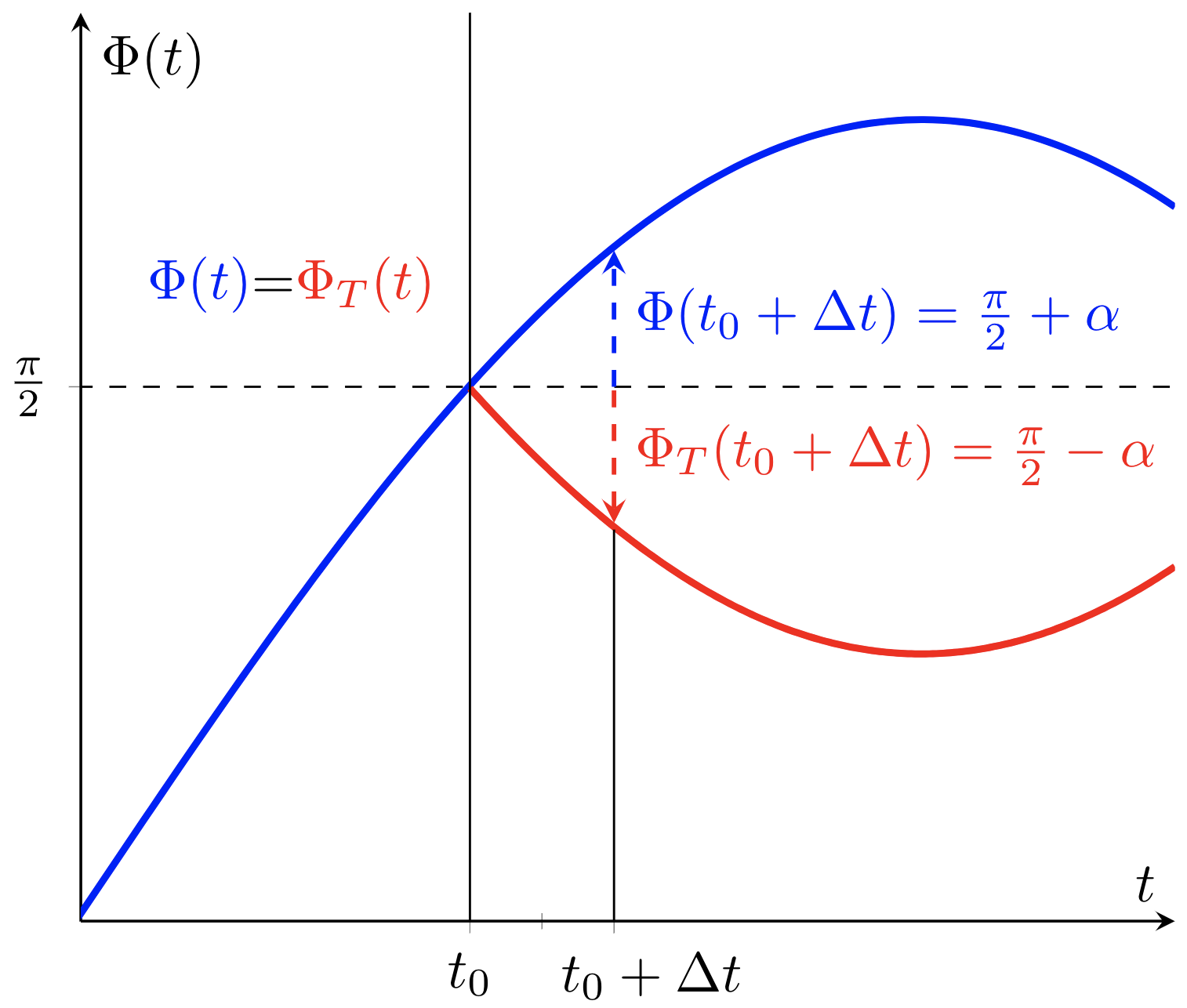}\end{center}\caption{When reproducing a current, while $\abs{\Phi(t)}<\frac{\pi}{2}$, the solution is unique and $\Phi(t)=\Phi_T(t)$. If however at some time $t_0$, $\Phi(t_0)<\frac{\pi}{2}$ and at the next time step $\Phi(t_0+\Delta t)=\frac{\pi}{2}+\alpha$, then the solution $\Phi_T(t_0+\Delta t)=\frac{\pi}{2}-\alpha$ will generate the same current, but ${\rm e}^{\pm  i\Phi_T(t_0+\Delta t)}={\rm e}^{\pm 2i\alpha}{\rm e}^{\pm i\Phi(t_0+\Delta t)}$, breaking the dynamical symmetry between the two systems.}
\label{fig:phiillustration}
\end{figure}
This phenomenon is illustrated in Fig.\ref{fig:phiillustration}, where crossing the threshold produces two solutions that will track the target observable. It is therefore possible to generate tracking control fields which reproduce the target, but have quite different dynamics, and hence, multiple solutions for  $\ket{\psi}$ and $\Phi_T(t)$.

We conclude this section with the observation that even in the case that $\ket{\psi}$ is unique, the tracking field $\Phi_T(t)$ defined in Eq.\eqref{eq:phi_track} will only be unique modulo $2\pi$. This constitutes a non-uniqueness in the tracking field \emph{at each timestep}. Fortunately, one is able to appeal to another physical principle to eliminate this non-uniqueness, namely that the system obey an \emph{Ehrenfest theorem} for current.

\section{Ehrenfest Theorems \label{sec:Ehrenfest}}
We now turn our attention to the question of verification of numerical simulations. Given the tracking strategy will by definition reproduce the trajectory one desires, it is important to have an independent check that tracking has been achieved via a physical evolution rather than numerical aberrations. A particularly sensitive test of the physicality of a numerical simulation is checking that expectations obey the relevant Ehrenfest theorems (see e.g. Ref.~\citep{PhysRevA.84.022326}). These relate derivatives of a given expectation to other expectations. In the Hubbard model, there is an Ehrenfest theorem for $J(t)$, namely
\begin{align}
   \frac{{\rm d} J\left(t\right)}{{\rm d}t}=&eat_{0}{\rm e}^{-i\Phi\left(t\right)} \sum_{j,\sigma}\left(\left<\left[\hat{H}\left(t\right),c_{j\sigma}^{\dagger}c_{j+1\sigma}\right]\right>\right. \nonumber \\ &\left.-\frac{{\rm d}\Phi(t)}{{\rm d}t}\left<c_{j,\sigma}^{\dagger}c_{j+1,\sigma}\right>\right) +\rm{h.c.} \label{eq:Ehrenfest theory}
\end{align}
which must be respected if the evolution is physical.

An important feature of the tracking Hamiltonian is that although the tracked variable will be reproduced by construction, there is no guarantee that any other observables will be tracked. This means that we only know \emph{a priori} the left hand side of Eq.\eqref{eq:Ehrenfest theory}, which will correspond by construction to $\frac{dJ_t}{dt}$, and can therefore verify a simulation respects physical principles by checking that the independent expectations from the right hand side of Eq.\eqref{eq:Ehrenfest theory} are correct. To do so, we assign the commutator in the first term of \eqref{eq:Ehrenfest theory} the following shorthand
\begin{align}
    \frac{1}{U}\sum_{j,\sigma}\left<\left[\hat{H}\left(t\right),c_{j\sigma}^{\dagger}c_{j+1\sigma}\right]\right>=C(\psi)\rm{e}^{i\kappa(\psi)}, \label{eq:twobodyexpectation}
\end{align}
from which we obtain an analytic expression for the current derivative in terms of the independent expectations defined by Eqs.(\ref{neighbourexpectation}) and (\ref{eq:twobodyexpectation}):
\begin{align}
\frac{{\rm d}J(t)}{{\rm d}t}=&-2eat_{0}\frac{{\rm d}\Phi(t)}{{\rm d}t}R\left(\psi\right)\cos\left(\Phi\left(t\right)-\theta\left(\psi\right)\right)\nonumber \\&-2eat_{0}U C(\psi)\cos\left(\Phi\left(t\right)-\kappa\left(\psi\right)\right),
\label{eq:Ehrenfest}
\end{align}
which provides a valuable consistency check for numerical simulations. 

The Ehrenfest theorem also resolves the problem of $\Phi_T(t)$ being only unique modulo $2\pi$ when $\ket{\psi}$ is unique. If at time $t, \Phi_T(t)$ correctly reproduces $J_T(t)$, then $\Phi_T(t)\to\Phi_T(t)+2n\pi$, $n\in\mathbb{Z}$  will generate the same current. This means that at each time, one in fact has an infinite number of choices for $\Phi_T(t)$. This non-uniqueness leads to $\Phi_T(t)$ being non differentiable. To see this consider
\begin{align}
\frac{{\rm d}\Phi(t)}{{\rm d}t}=\lim_{\Delta t\to0}\frac{\Phi_T(t+\Delta t)-\Phi_T(t)}{\Delta t}. \label{eq:philimit}
\end{align}
 If the derivative exists for this solution, then switching solution to (for instance) $\Phi_T(t+\Delta t)\to \Phi_T(t+\Delta t)+2n\pi$ would render $\Phi_T$ non-differentiable, as the limit on the right hand side of Eq.\eqref{eq:philimit} would not exist. For the Ehrenfest theorem to be meaningful however, $\frac{{\rm d}\Phi(t)}{{\rm d}t}$ must exist. For this reason, the additional solutions resulting from adding integer multiples of $2n\pi$ at any time cannot be admitted as physical. Eq.(\ref{eq:Ehrenfest})  uniquely specifies $\frac{{\rm d}\Phi(t)}{{\rm d}t}$, and stipulating that the evolution must obey this means that for a given initial condition, $\Phi_T(t)$ has a unique solution.
To test the Ehrenfest theorem, we take two systems at $U=0$  and $U=7t_0$, and drive them with the $\Phi(t)$ shown in Fig.\ref{fig:multiplesimulation}. All results are obtained with a numerically exact time propagation of the correlated state. More details for these reference systems can be found in Ref.\citep{companionletter}.

\begin{figure}
\begin{center}
\includegraphics[width=1.1\columnwidth]{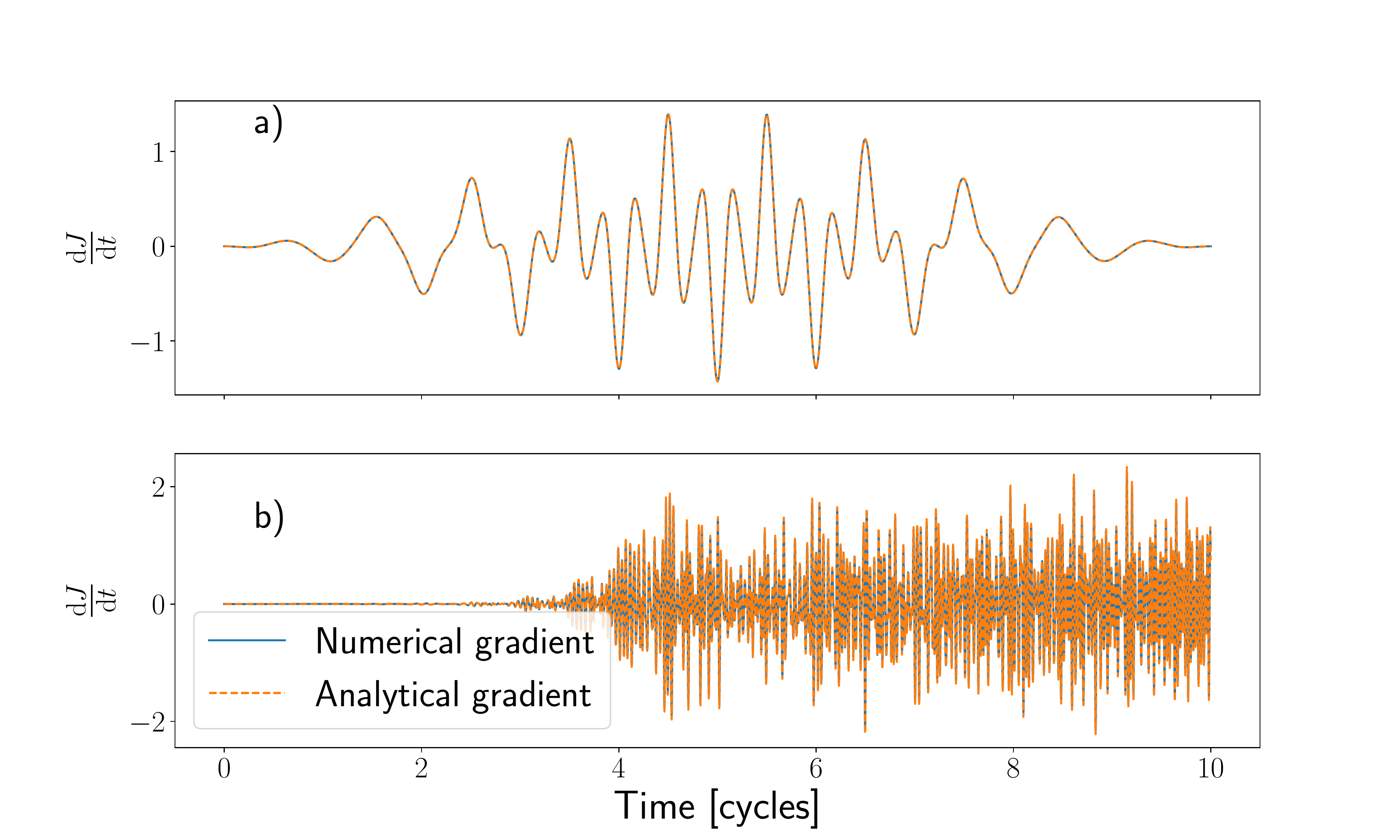}\end{center}\caption{Comparison between the numerical current gradient, and the analytic prediction calculated via Eq.(\ref{eq:Ehrenfest}) for both \textbf{a)} $\frac{U}{t_0}=0$ and \textbf{b)} $\frac{U}{t_0}=7$ when driven by the $\Phi(t)$ shown in Fig.\ref{fig:multiplesimulation})}
\label{fig:refEhrenfest}
\end{figure}

\begin{figure}
\includegraphics[width=1.1\columnwidth]{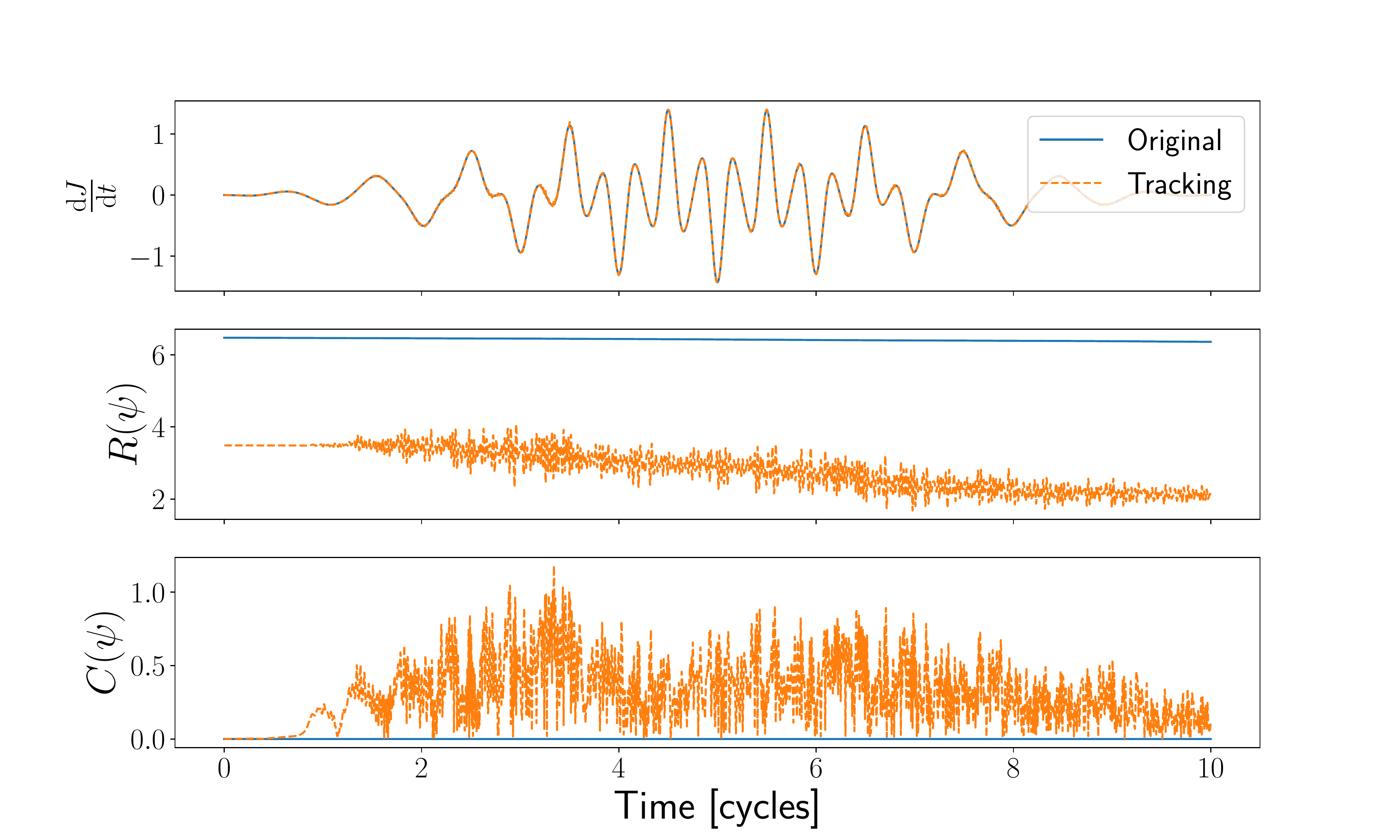}\caption{When tracking the original $J(t)$ from the $U=0$ system in the $U=7t_0$ system, we find that  $\frac{{\rm d}J_T}{{\rm d}t}$ when calculated via Eq.(\ref{eq:Ehrenfest}) agrees with the numerical gradient (top panel). This is despite the fact that the untracked expectations $R(\psi)$ and $C(\psi)$ used in Eq.\eqref{eq:Ehrenfest} have different trajectories for each of the two simulations.}
\label{fig:Ehrenfesttracking}
\end{figure}

Fig.\ref{fig:refEhrenfest} compares the dipole acceleration $\frac{{\rm d} J(t)}{{\rm d}t}$ calculated using Eq.(\ref{eq:Ehrenfest}) to the numerical gradient. It can be seen from that both calculations align perfectly, as they must for the system evolution to be considered physical. Extending this to tracking control, Fig.\ref{fig:Ehrenfesttracking}  provides an example demonstrating that the Ehrenfest theorem is obeyed when tracking the current of a different system. This highlights the fact that the theorem is obeyed in two systems with the same current gradient, despite the fact that the non-tracked expectations do not match between simulations.

\begin{figure}
\includegraphics[width=1\columnwidth]{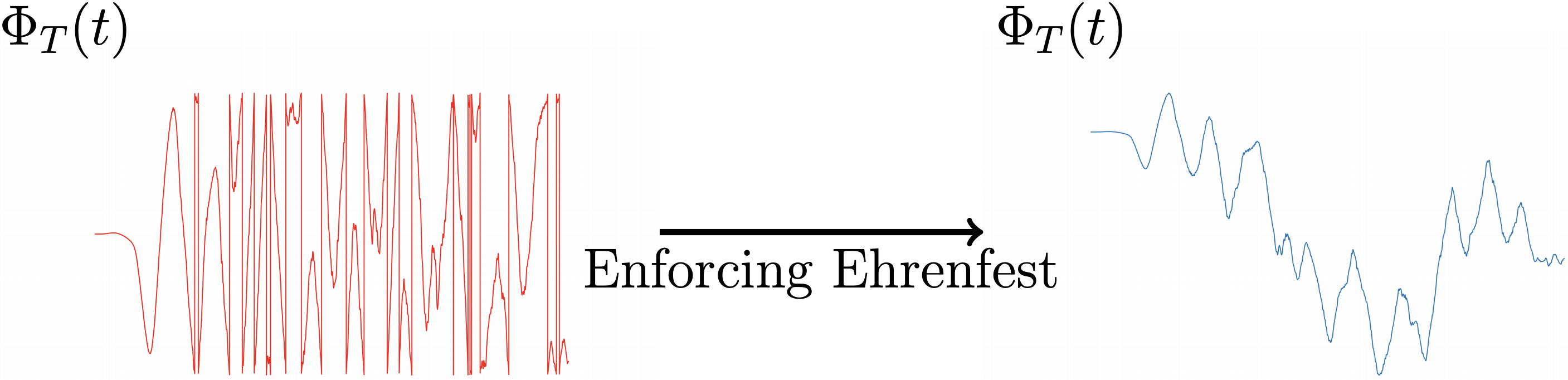}\caption{While $\Phi(t)-\theta(\psi)$ is always constrained to lie between $\left[-\frac{\pi}{2},\frac{\pi}{2} \right]$ modulo $2\pi$, both $\theta(\psi)$ and $\Phi(t)$ can individually undergo large oscillations which introduce numerical discontinuities. These unphysical discontinuities can be identified by appealing to the Ehrenfest theorem, and removed so as to enforce Eq.\eqref{eq:Ehrenfest theory}.} 
\label{fig:Ehrenfestunwravel}
\end{figure}

  The verification provided by Ehrenfest theorems is particularly useful for tracking in high $\frac{U}{t_0}$ simulations, when $\theta(\psi)$ exhibits large oscillations. When this angle is calculated numerically, it is given a value between $\left[-\pi,\pi \right]$. If on a timestep update this threshold is crossed, a numerical discontinuity is introduced by the assignment $\theta(\psi)=\pm \pi \pm \delta \to \mp \pi \pm \delta$. The Ehrenfest theorem is sensitive to this artifical discontinuity, and can therefore be used to correct it in both $\theta(\psi)$ and $\Phi_T(t)$. An example of a control field where this correction is necessary is shown in Fig.\ref{fig:Ehrenfestunwravel} 
 
\section{Experimental Feasibility \label{sec:Experimental}}
Although the previous section, and the material mimicry done in Ref.\citep{companionletter} demonstrates that the tracking strategy is successful \emph{in silico}, there remains a question of the experimental feasibility of generating the laser pulses prescribed by the tracking strategy. Although it is possible to implement a control scheme which reflects experimental constraints \citep{Kosloffoptimalconstraints, SpectralConstraints}, this in general does not guarantee an exact match with the target. In order to guarantee exact tracking in Ref. \citep{companionletter}, neither the intensity nor bandwidth of the driving field was constrained.

As a first test of the experimental feasibility of our method, we examine the effect of introducing a cut-off frequency $\omega_c$ to the control field obtained from the material mimicry in Ref.\citep{companionletter}. Taking $\Phi_T(t)$ from Eq.\eqref{eq:phi_track}, we make a cut-off in frequency space such that ${\widetilde{\Phi}_T(\omega>\omega_c)=0}$. This post-processed control field is then used to solve the Schr{\"o}dinger equation for the same system $\Phi_T(t)$ was originally applied to. 
\begin{figure}
\begin{center}
\includegraphics[width=1.1\columnwidth]{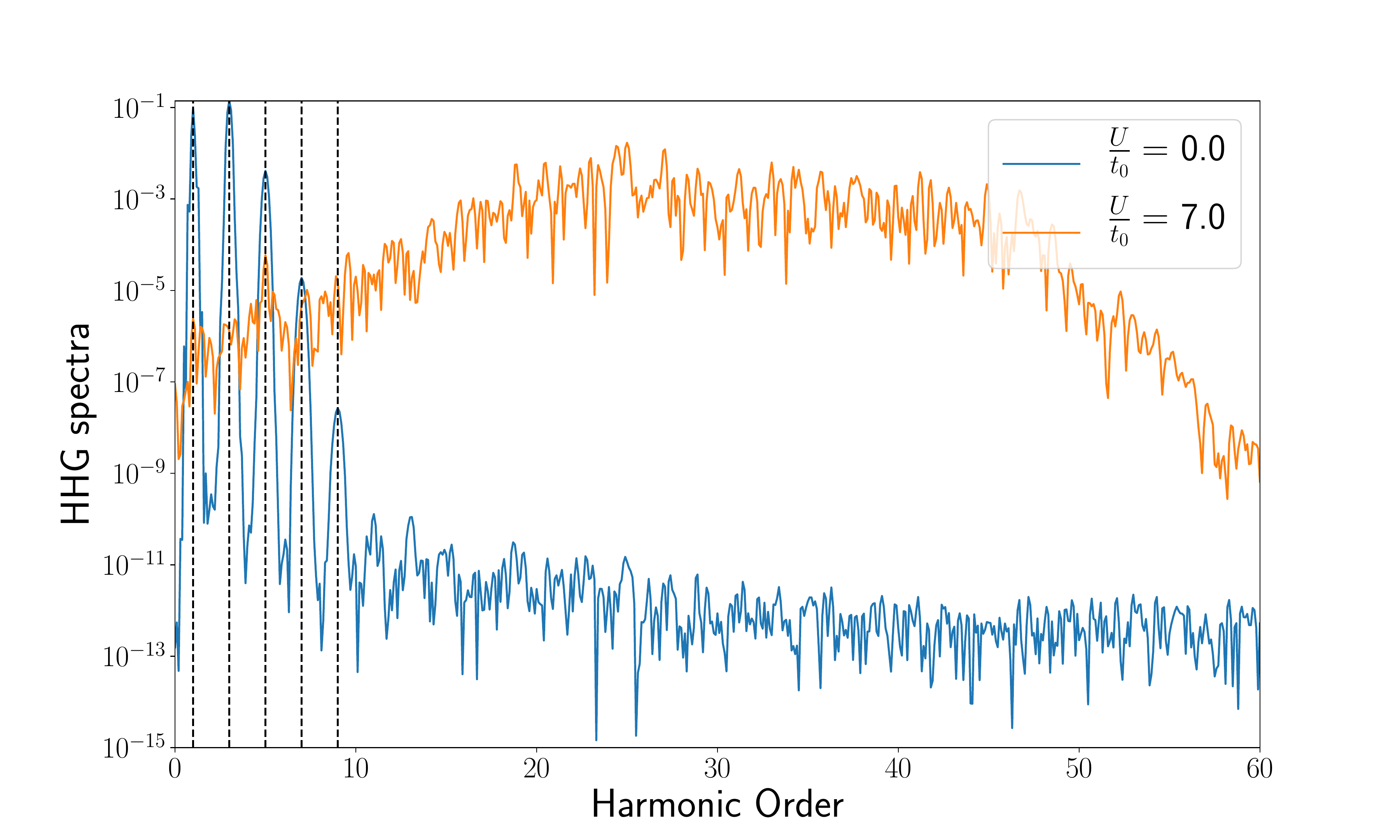}\end{center}\caption{Reference harmonic spectra for both the conducting limit $\frac{U}{t_0}=0$ and insulating  $\frac{U}{t_0}=7$ states, generated using the $\Phi(t)$ shown in Fig.\ref{fig:multiplesimulation}. Dashed vertical lines indicate odd order overtones, where single-band models predict harmonic generation.}
\label{fig:refspectra}
\end{figure}

\begin{figure}
\includegraphics[width=1.1\columnwidth]{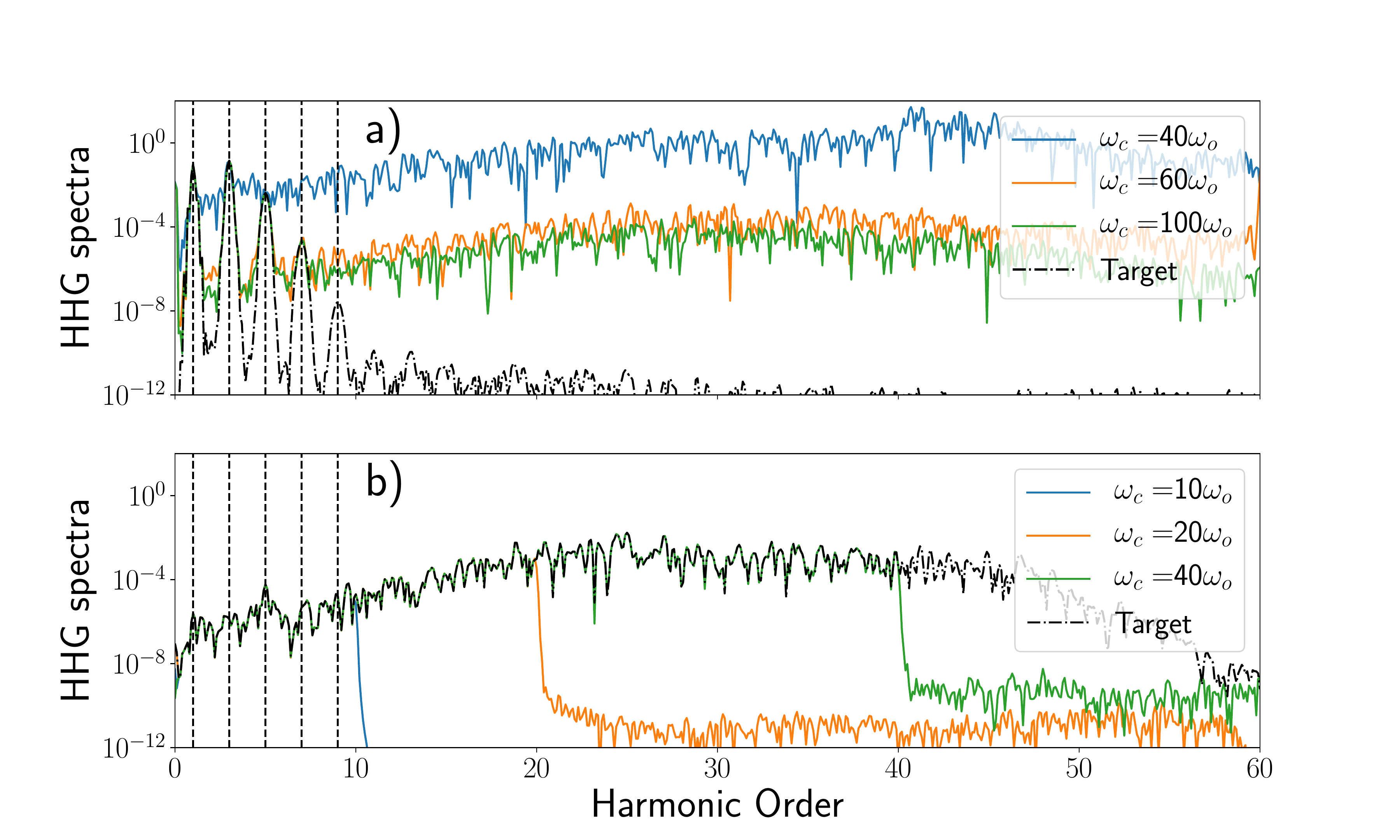}\caption{Introducing a low-pass filter on $\Phi_T(t)$ before using it to evolve the system, we find that for \textbf{a)} $U=7t_0$ tracking the $U=0$ reference spectrum, a rather high cut-off frequency $\omega_c$ is required to track the first four most prominent harmonics of the target spectrum, while in the converse case \textbf{b)},  a linear dependence to $\omega_c$ is observed}
\label{fig:J7cut}
\end{figure}

As targets, we use reference spectra (the Fourier transform of the dipole accelerations presented in Fig.\ref{fig:refEhrenfest}), which are shown in Fig.\ref{fig:refspectra}, and results can be seen in Fig.\ref{fig:J7cut}. Two conclusions can be drawn from these results. First, when tracking the insulating systems spectrum in the conducting limit, as shown in Fig.\ref{fig:J7cut}\textbf{(b)}, the spectra matches its target well while $\omega<\omega_c$, after which it is strongly suppressed. Conversely, in the case where a system with very strong onsite-repulsion tracks the tight-binding spectrum (as in Fig.\ref{fig:J7cut}\textbf{(a)}), the response to the cut-off appears highly non-linear, and a very broadband pulse with a cutoff of $\omega_c \approx 50\omega_0$ is needed to reproduce the four most prominent harmonics associated with $J^{(0)}(t)$. This suggests that when two materials are at greater distances from each other in the phase diagram, greater bandwidth in the control field is required for tracking.

\begin{figure}
\includegraphics[width=1.1\columnwidth]{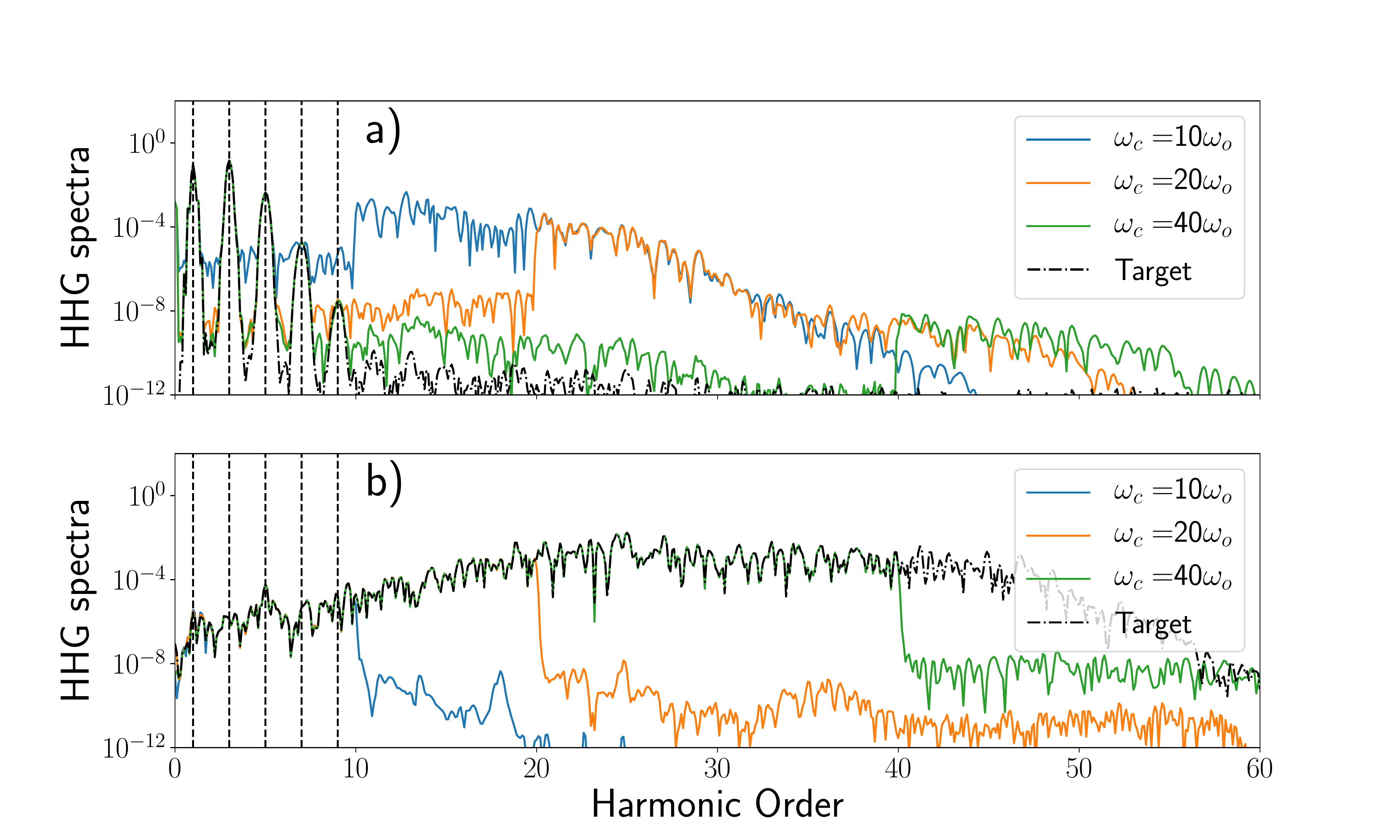}\caption{Tracking both reference spectra at $\frac{U}{t_0}=1$, one finds that for  both \textbf{a)} $U=0$ and \textbf{b)} $U=7$ reference spectra, a linear dependence to the cut-off $\omega_c$ is observed. When tracking $J^{(0)}(t)$ in this intermediate system, the main features of the target spectrum are captured at much lower cut-off frequencies, as compared to the results in Fig.\ref{fig:J7cut}.}
\label{fig:J1cut}
\end{figure}

When tracking both reference spectra in an intermediate material $\frac{U}{t_0}=1$, we find more promising results, as shown in Fig.\ref{fig:J1cut}. In this case, a linear dependence on $\omega_c$ is observed in both tracked spectra, and one is able to recover the most prominent harmonics of the $\frac{U}{t_0}=0$ reference system at a potentially realisable $\omega_c=10\omega_0$.

\section{Discussion \label{sec:Discussion}}
In this paper we have expanded on the work presented in Ref.\citep{companionletter}. In addition to providing a more complete derivation for the tracking model's equation of motion, constraints guaranteeing Hermiticity and a unique evolution were rigorously derived. Although these constraints restrict the size of imitable currents in tracking, this can be circumvented either by scaling the current one wishes to track, or modifying system parameters such that the constraints are obeyed. The ability to transparently identify and remove singularities via scaling represents a tangible advantage over more generic tracking strategies \citep{doi:10.1063/1.1582847, doi:10.1137/0325030, PhysRevA.98.043429}. 

The derived constraints of Eqs.(\ref{eq:constraint2},\ref{eq:constraint1}) also highlight an interesting ambiguity in the tracking model, namely that in some circumstances multiple control fields will track the same target expectation. This raises a question for future investigations about the enumeration of these solutions, and how their dynamics differ.  
An Ehrenfest theorem for the tracked expectation was also introduced for the purpose of verifying the consistency of the numerics with the constraints of physical principles. By insisting that this Ehrenfest theorem be obeyed removes unphysical discontinuities that can arise from the periodic effect of $\Phi(t)$ on the dynamics.

In investigating the potential to realize this tracking experimentally with finite-bandwidth applied fields, we employed a low-pass filter on the tracking control field $\Phi_T(t)$. This produced an interesting asymmetry in the tracking response to the cut-off frequency $\omega_c$. While systems tracking currents generated by materials with a \emph{higher} $\frac{U}{t_0}$ always displayed a linear dependence, this linearity was not always observed for the converse case of tracking \emph{lower} $\frac{U}{t_0}$. While there appears to be a regime of linear dependence when the gap between the original and tracked system parameters is sufficiently small (see Fig.\ref{fig:J1cut}), Fig.\ref{fig:J7cut} shows that when the two systems are separated by greater distances on the phase diagram, the current response to a cut-off in $\Phi_T(t)$ is highly non-linear. 

To achieve the fine control over expectations shown both in this paper and Ref.\citep{companionletter}, it will be necessary to adapt the tracking strategy to reflect experimental constraints. A potential future avenue is to optimise tracking results while only utilising a small number of discrete, experimentally feasible frequencies, rather than the unrestricted broadband pulses used in the simulations presented here.  
Finally, the same concepts used to derive the model presented here could potentially be applied to optimal dynamic discrimination (ODD). This problem is essentially the converse to that of tracking control, in which one distinguishes very similar quantum systems using the dynamics induced by properly shaped laser pulses \citep{oddrabitz,Goun_Bondar_Er_Quine_Rabitz_2016}. Given that the requirements for discrimination are similar to those for tracking control, the former may benefit from the techniques presented here.

\section{Acknowledgements}
G.M. and D.I.B. are supported by Air Force Office of Scientific Research (AFOSR) Young Investigator Research Program (grant FA9550-16-1-0254) and the Army Research Office (ARO) (grant W911NF-19-1-0377). The views and conclusions contained in this document are those of the authors and should not be interpreted as representing the official policies, either expressed or implied, of AFOSR, ARO, or the U.S. Government. The U.S. Government is authorized to reproduce and distribute reprints for Government purposes notwithstanding any copyright notation herein.

G.H.B and C.O. acknowledge funding by the Engineering and Physical Sciences Research Council (EPSRC) through the Centre for Doctoral Training ``Cross Disciplinary Approaches to Non-Equilibrium Systems" (CANES, Grant No. EP/L015854/1). G.H.B. gratefully acknowledges support from the Royal Society via a University Research Fellowship, and funding from the Air Force Office of Scientific Research via grant number FA9550-18-1-0515. The project has received funding from the European Union's Horizon 2020 research and innovation programme under grant agreement No. 759063.

\end{document}